\documentclass[aps,prb,twocolumn,superscriptaddress,groupeaddress]{revtex4}

\usepackage{graphicx}% Include figure files
\usepackage{dcolumn}% Align table columns on decimal point

\begin{document}

\title{
Flux-driven Josephson parametric amplifier
}

\author{T. Yamamoto}
\affiliation{NEC Nano Electronics Research Laboratories, Tsukuba, Ibaraki 305-8501, Japan}
\affiliation{Emergent Materials Department, The Institute of Physical and Chemical Research (RIKEN), Wako-shi, Saitama 351-0198, Japan} 
\affiliation{CREST-JST, Kawaguchi, Saitama 332-0012, Japan} 

\author{K. Inomata}
\affiliation{Emergent Materials Department, The Institute of Physical and Chemical Research (RIKEN), Wako-shi, Saitama 351-0198, Japan} 

\author{M. Watanabe}
\affiliation{Emergent Materials Department, The Institute of Physical and Chemical Research (RIKEN), Wako-shi, Saitama 351-0198, Japan} 

\author{K. Matsuba}
\affiliation{CREST-JST, Kawaguchi, Saitama 332-0012, Japan} 
\affiliation{Department of Materials Sciences and Engineering, Tokyo Institute of Technology, 4259 Nagatsuta-cho, Midori-ku, Yokohama, 226-8503, Japan} 

\author{T. Miyazaki}
\affiliation{CREST-JST, Kawaguchi, Saitama 332-0012, Japan} 

\author{W. D. Oliver}
\affiliation{MIT Lincoln Laboratory, 244 Wood Street, Lexington, Massachusetts 02420, USA} 

\author{Y. Nakamura}
\affiliation{NEC Nano Electronics Research Laboratories, Tsukuba, Ibaraki 305-8501, Japan}
\affiliation{Emergent Materials Department, The Institute of Physical and Chemical Research (RIKEN), Wako-shi, Saitama 351-0198, Japan} 
\affiliation{CREST-JST, Kawaguchi, Saitama 332-0012, Japan} 

\author{J. S. Tsai}
\affiliation{NEC Nano Electronics Research Laboratories, Tsukuba, Ibaraki 305-8501, Japan}
\affiliation{Emergent Materials Department, The Institute of Physical and Chemical Research (RIKEN), Wako-shi, Saitama 351-0198, Japan} 
\affiliation{CREST-JST, Kawaguchi, Saitama 332-0012, Japan}

\date{\today}

\begin{abstract}
We have developed a Josephson parametric amplifier, 
comprising a superconducting coplanar waveguide resonator terminated by 
a dc SQUID (superconducting quantum interference device). 
An external field (the pump, $\sim 20$ GHz) modulates the flux threading the dc SQUID, 
and, thereby, the resonant frequency of the cavity field (the signal, $\sim 10$ GHz), 
which leads to parametric signal amplification. 
We operated the amplifier at different band centers, 
and observed amplification (17 dB at maximum) and deamplification 
depending on the relative phase between the pump and the signal. 
The noise temperature is estimated to be less than 0.87 K. 

\end{abstract}

%\pacs{
%\keywords{}

\maketitle

Degenerate parametric amplifiers are phase sensitive amplifiers, 
which can in principle amplify one of the two quadratures of 
a signal without introducing extra noise.~\cite{Takahashi65,Caves82} 
Parametric amplifiers based on the nonlinear inductance of a Josephson junction 
have been studied for a long time,~\cite{Baronebook} 
including one which demonstrated vacuum noise squeezing.~\cite{Movshovich90} 
Recently, there has been a renewed interest in parametric amplifiers~\cite{Abdo06,Tholen07,Beltran07} 
due in part to the increasing need for quantum-limited amplification 
in the field of quantum information processing using 
superconducting circuits.~\cite{Nakamura99,Wallraff04} 

In the present work, we design a Josephson parametric amplifier, 
comprising a superconducting transmission-line resonator terminated by a dc SQUID 
(superconducting quantum interference device). 
Contrary to the previous works, the pump is not used to directly modulate a current through the Josephson junction, 
but is instead used to modulate a flux through the dc SQUID.~\cite{Ojanen07} 
The resonant frequency of the resonator, namely, the band center of the signal, is widely controllable by 
a dc flux also applied to the SQUID ~(see also Ref.~[\onlinecite{Beltran07}]). 
Moreover, as the pump and the signal are applied to different ports and 
their frequencies are twice different (see below), 
it is straightforward to separate the output signal from the pump. 
This is a unique property of the flux-pumping scheme; 
it arises because the  finite dc flux bias allows linear coupling of the pump even in the absence of a dc current bias 
across the SQUID.~\cite{comment1}
%\footnote{
%In the current-pumping scheme, the pump modulates Josephson inductance only quadratically 
%unless the SQUID is biased by a dc current. Therefore, in the degenerate-mode operation, 
%the frequencies of the pump and the signal are equal.\cite{Beltran07}
%}
We operated such a flux-driven parametric amplifier and characterized its basic properties. 

\begin{figure}
\includegraphics[width=0.9\columnwidth,clip]{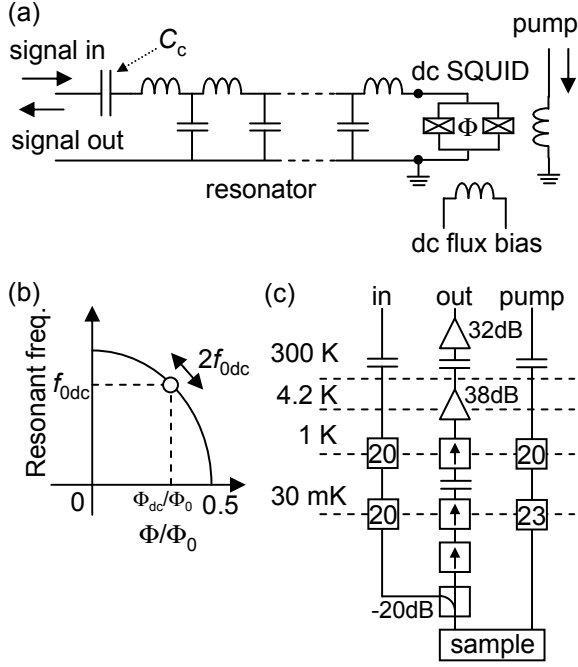}
\caption{\label{fig:fig1}
(a) Schematic circuit diagram of the flux-driven Josephson parametric amplifier. 
(b) Operation principle of the parametric amplifier. Solid curve represents the resonant 
frequency of the cavity. Ac flux modulation induces the modulation of the resonant 
frequency about its static value $f_{0{\rm dc}}$, which leads to the parametric amplification. 
(c) Schematic diagram of the measurement setup. Squares with a number inside represent 
fixed attenuators with corresponding attenuation in dB. 
Squares with an arrow inside represent isolators. 
}
\end{figure}

\begin{figure}
\includegraphics[width=0.9\columnwidth,clip]{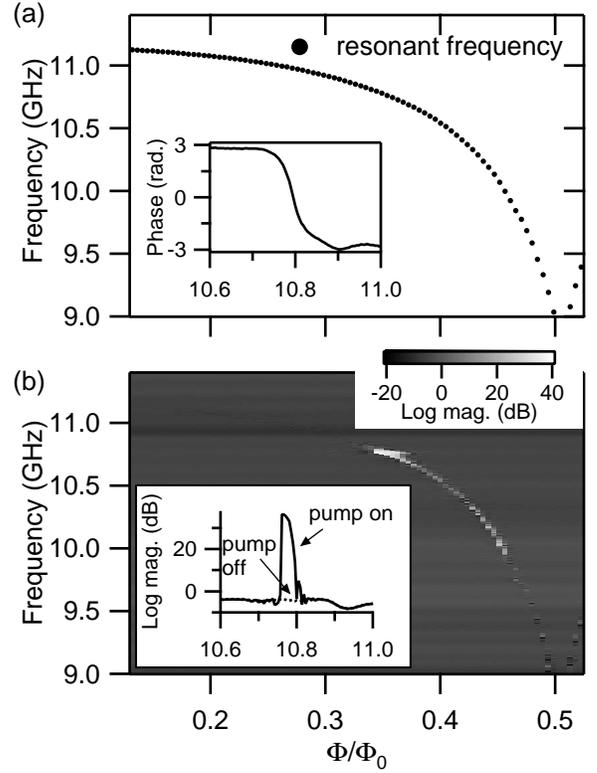}
\caption{\label{fig:fig2}
(a) Resonant frequency as a function of the normalized magnetic flux. 
The inset shows the phase of the reflection coefficient $\Gamma$ as a function of the 
frequency in GHz at $\Phi/\Phi_0=0.34$. 
(b) An intensity plot of $|\Gamma|$ (pump on) 
as a function of the signal 
frequency and the normalized magnetic flux. 
The inset shows a cross section at $\Phi/\Phi_0=0.34$ together with 
the data when the pump is off. 
}
\end{figure}

Figure~\ref{fig:fig1}a shows a schematic diagram of the flux-driven parametric amplifier. 
The primary component of the amplifier is a transmission-line resonator defined by its coupling 
capacitance $C_{\rm c}$ and a dc SQUID termination. 
The magnetic flux $\Phi$ penetrating the SQUID loop changes 
the boundary condition of the resonator at the right end (by the change of the Josephson inductance), 
and hence enables us to control the resonant frequency.~\cite{Wallquist06,Sandberg08} 
The resonant frequency $f_0$ for the first mode ($\lambda/4 \ge d$, where $\lambda$ is the 
wavelength and $d$ is the cavity length) 
is schematically drawn as a function of $\Phi/\Phi_0$ in Fig.~\ref{fig:fig1}b, 
where $\Phi_0$ is a flux quantum (see also Fig.~\ref{fig:fig2}a). 
We now assume the cavity resonance is set to a particular value, $f_{0{\rm dc}}$, 
by applying a dc flux $\Phi_{\rm dc}$ (open circle in the figure). 
We then apply microwaves at a frequency $2f_{0{\rm dc}}$ to the pump line 
which is inductively coupled to the SQUID loop. 
The pump microwaves modulate $\Phi$ and, hence, $f_0$ 
about its static value $f_{0{\rm dc}}$ at $2f_{0{\rm dc}}$. 
This modulation does parametric work for the signal microwaves at $f_{0{\rm dc}}$ 
which comes into the resonator. The amplified (or deamplified) signal is then reflected 
back along the signal line. 
Note that the leakage of the pump microwaves into the signal port is suppressed, 
because the second resonance mode does not exist around $2f_{0{\rm dc}}$. 

%%%%%%%%%%%%%%%%%%%%%%%%%%%%%%%%%%%%%%%%%%%%%%%%%
%%%%%  Fig. 1 = fig1  
%%%%%%%%%%%%%%%%%%%%%%%%%%%%%%%%%%%%%%%%%%%%%%%%%
%%%%%%%%%%%%%%%%%%%%%%%%%%%%%%%%%%%%%%%%%%%%

The device was fabricated using the planarized niobium trilayer process at MIT Lincoln Laboratory.
The coplanar-waveguide resonator is made by patterning a 150-nm-thick niobium 
film deposited on an oxidized silicon substrate. It consists of a center conductor (10~$\mu$m wide and 2.6 mm long)
and two lateral ground planes nearby. 
The coupling capacitance $C_{\rm c}$ was designed to be 16 fF. 
The critical current of the Josephson junctions for the SQUID was estimated to be 
1.5~$\mu$A per each from the process test data. 
The self-inductance of the SQUID loop and the mutual inductance between the SQUID loop and the pump line 
were estimated by a numerical calculation to be 34 and 5 pH, respectively. 
The field for the dc flux bias was generated by a superconducting solenoid, 
which is attached below the sample package. 

All measurements were done using a dilution refrigerator at the base temperature of 30 mK. 
Figure~\ref{fig:fig1}c is a schematic diagram of the measurement setup. 
We used a directional coupler, which gave 20-dB attenuation to the input signal, 
to measure the reflected microwaves from the resonator.  
We also used a cryogenic high electron mobility transistor (HEMT) 
amplifier with a gain of 38 dB and a noise temperature $T_{\rm N}^{{\rm HEMT}}$ of 9~K at 10 GHz. 
The frequency band for the signal line, which is limited by the isolators, is roughly from 9 to 11 GHz. 

First, we characterized the resonator using a vector network analyzer 
connected to the ``in'' and  ``out'' ports in Fig.~\ref{fig:fig1}c. 
All power levels shown below refer to the input or the output of the sample box, 
where gains or losses of the components (e.g. HEMT amplifiers and attenuators) 
in the measurement lines have been calibrated away within an accuracy of $\pm2$ dB. 
Under a fixed dc flux bias, we observe a 2$\pi$ rotation of the phase of the reflection coefficient $\Gamma$, 
which is measured as the scattering parameter $S_{21}$, at the resonant frequency. 
The inset of Fig.~\ref{fig:fig2}a shows an example for $\Phi/\Phi_0=0.34$, 
from which the quality factor of the resonator is estimated to be 250.~\cite{comment2}
%\footnote{The quality factor is flux dependent. It tends to be smaller as we approach $\Phi/\Phi_0=0.5$. 
%For example, it is 170 for $\Phi/\Phi_0=0.42$.}
Here the power of the input signal $P_{\rm s}$ is sufficiently low (-136 dBm). 

The resonant frequency $f_{0{\rm dc}}$ exhibits a periodic dependence on $\Phi/\Phi_0$. 
It becomes maximal (11.16 GHz) at $\Phi/\Phi_0=m$ and decreases 
as we approach $\Phi/\Phi_0=m+1/2$, where $m$ is an integer (see Fig.~\ref{fig:fig1}b). 
A part of the modulation is shown in Fig.~\ref{fig:fig2}a. 
The minimum is not exactly at 0.5 because of the nonzero loop inductance of the dc SQUID. 

%%%%%%%%%%%%%%%%%%%%%%%%%%%%%%%%%%%%%%%%%%%%%%%%%
%%%%%  Fig. 2 = fig2  
%%%%%%%%%%%%%%%%%%%%%%%%%%%%%%%%%%%%%%%%%%%%%%%%%
%%%%%%%%%%%%%%%%%%%%%%%%%%%%%%%%%%%%%%%%%%%%

Next, we performed similar measurements in the presence of the pump microwaves applied to the ``pump'' port. 
Figure~\ref{fig:fig2}b is an intensity plot of the $\Gamma$ amplitude 
as a function of the signal frequency $f_{\rm s}$ and $\Phi/\Phi_0$. 
As $f_{\rm s}$ is swept by the network analyzer, 
the pump frequency $f_{\rm p}$, which is generated by another microwave source, 
is synchronously swept such that $f_{\rm p}=2f_{\rm s}$. 
The power of the pump microwaves $P_{\rm p}$ is -66 dBm. 
We observe amplified signals near the corresponding resonant frequencies. 
The bright spot near $\Phi/\Phi_0=0.34$ is observed even without input signal 
when $P_{\rm p}$ is larger than -72 dBm. 
As shown in the inset, which is a cross section at $\Phi/\Phi_0=0.34$, 
the gain looks much larger than those shown later. 
Here, the device is working as an oscillator rather than an amplifier. 

\begin{figure}
\includegraphics[width=0.9\columnwidth,clip]{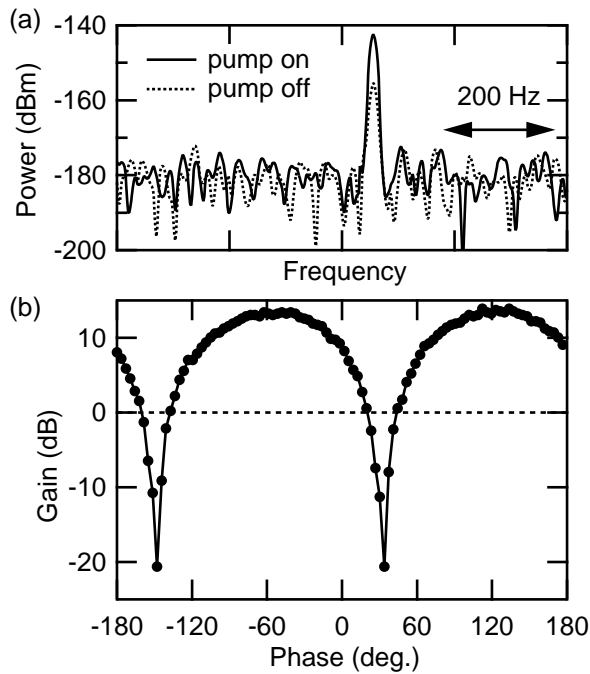}
\caption{\label{fig:fig3}
(a) Side-band peak measured by the spectrum analyzer. 
The peak is located at 10.78 GHz + 10 kHz (carrier + modulation). 
The solid curve is a spectrum when the pump is on, while the 
dashed curve is a spectrum when the pump is off. 
(b) Gain as a function of the carrier phase of the signal. 
}
\end{figure}

To further characterize the amplifier, we studied the phase dependence of the gain, 
a hallmark of degenerate parametric amplifiers, using two phase-locked microwave generators. 
The signal at $f_{\rm s}$ was amplitude modulated at 10 kHz and applied to the signal input. 
The pump at $f_{\rm p}=2f_{\rm s}$ (twice the carrier of the input signal) was applied to the pump port. 
A spectrum analyzer was used to detect the output power at the side-band frequency ($f_{\rm s}+10$ kHz). 
Figure~\ref{fig:fig3}a shows the side-band peak measured by the spectrum analyzer 
with 10-Hz resolution bandwidth and video bandwidth. 
For comparison, the trace when the pump is turned off is also plotted. 
Here, $f_{0{\rm dc}}=10.77$ GHz, $f_{\rm s}=10.78$ GHz, $f_{\rm p}=21.56$ GHz, $P_{\rm s}=-148$ dBm, 
and $P_{\rm p}=-71$ dBm. 
We intentionally detuned $f_{\rm s}$ from $f_{0{\rm dc}}$ 
to avoid having emission without an input field, as was mentioned earlier. 
We defined the gain as the difference between the two side-band peak values and 
studied how it depends on the relative phase between the carriers of the signal and the pump. 

Figure~\ref{fig:fig3}b shows the gain as a function of the carrier phase of the signal. 
As expected, it shows a periodic dependence with a period of $\pi$. 
The maximum gain is 14 dB.~\cite{comment3} 
At certain phases, the peak value becomes lower than the pump-off level, and the gain is negative. 
This is the so-called deamplification, a key property associated with noise squeezing.~\cite{Yurke88,Movshovich90} 

%%%%%%%%%%%%%%%%%%%%%%%%%%%%%%%%%%%%%%%%%%%%%%%%%
%%%%%  Fig. 3 = fig3  
%%%%%%%%%%%%%%%%%%%%%%%%%%%%%%%%%%%%%%%%%%%%%%%%%
%%%%%%%%%%%%%%%%%%%%%%%%%%%%%%%%%%%%%%%%%%%%

\begin{figure}
\includegraphics[width=0.9\columnwidth,clip]{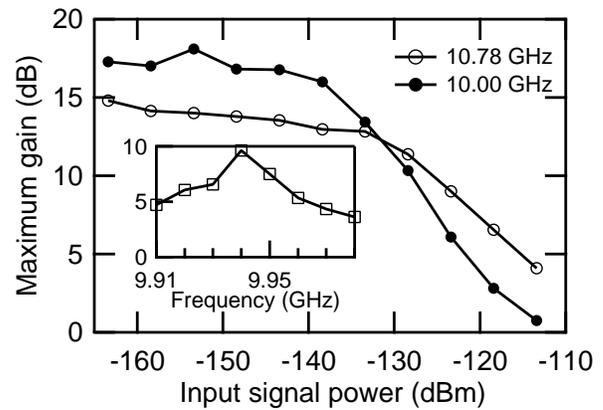}
\caption{\label{fig:fig4}
Maximum gain as a function of the input signal power $P_{\rm s}$. 
For $f_{\rm s}=10.00$~GHz and 10.78~GHz, $f_{0{\rm dc}}=10.00$~GHz and 10.77~GHz, respectively. 
The inset shows a maximum gain as a function of the input signal frequency 
for $f_{0{\rm dc}}=9.94$ GHz. 
}
\end{figure}

Figure~\ref{fig:fig4} shows the maximum value of the phase-dependent gain as a function of $P_{\rm s}$. 
In the case of $f_{\rm s}=10.78$ GHz, the maximum gain decreases quickly at $P_{\rm s}>-130$ dBm. 
The lower bound is not determined by the parametric amplifier itself, but by the noise of the measurement setup, 
which is dominated by the noise of the HEMT amplifier. 
The gain suppression at high $P_{\rm s}$ is closely related to the nonlinear inductance of the Josephson junctions. 
Without pumping, we observe a hysteretic behavior (bifurcation) 
in the frequency dependence of $\Gamma$ at above $P_{\rm s}=-101$ dBm. 
Also, at high $P_{\rm s}$, we observe a distorted (sawtooth-like) curve in the 
phase dependence of the gain, which becomes conspicuous above $P_{\rm s}\sim-120$ dBm (data not shown). 

As we mentioned, one of the advantages of the present parametric amplifier is 
the controllability of the band center. 
To demonstrate this, we adjusted $\Phi_{\rm dc}$ so that $f_{0{\rm dc}}=10.00$ GHz and 
operated the amplifier at $f_{\rm s}=10.00$ GHz, $f_{\rm p}=20.00$ GHz, and $P_{\rm p}=-71$ dBm.
Qualitatively the same phase dependence was observed, and the maximum gain 
as a function of $P_{\rm s}$ is plotted in Fig.~\ref{fig:fig4}. 
Although the dynamic range is narrower because of a smaller critical current of the dc SQUID, 
a gain of 17 dB is attained. 

A bandwidth at fixed $f_{0{\rm dc}}$ was also investigated, as 
exemplified in the inset of Fig.~\ref{fig:fig4}. 
Here, $f_{0{\rm dc}}$ is set to be 9.94 GHz, and $f_{\rm s}$ was varied around $f_{0{\rm dc}}$. 
For each $f_{\rm s}$, we performed a phase-dependence measurement to determine the maximum gain, 
and plot it as a function of $f_{\rm s}$. 
We set $f_{\rm p}=2f_{\rm s}$, $P_{\rm s}=-143$ dBm, and $P_{\rm p}=-71$ dBm. 
As seen from this plot, the bandwidth of the parametric amplifier at fixed $f_{0{\rm dc}}$ is about 20 MHz. 
This value does not depend much on $f_{0{\rm dc}}$, and is comparable to the linewidth of the cavity. 

%%%%%%%%%%%%%%%%%%%%%%%%%%%%%%%%%%%%%%%%%%%%%%%%%
%%%%%  Fig. 4 = fig4  
%%%%%%%%%%%%%%%%%%%%%%%%%%%%%%%%%%%%%%%%%%%%%%%%%
%%%%%%%%%%%%%%%%%%%%%%%%%%%%%%%%%%%%%%%%%%%%

The noise temperature $T_{\rm N}$ is another important figure of merit. 
In Fig.~\ref{fig:fig3}a, the pump increases the signal level by 14 dB, but 
hardly changes the background noise level, which is limited by $T_{\rm N}^{{\rm HEMT}}$. 
This implies that $T_{\rm N}^{{\rm HEMT}}>(T_{\rm Ni}+T_{\rm N})G_{\rm nd}/L$, 
where $T_{\rm Ni}$ is the input noise temperature, 
$G_{\rm nd}$ is the gain of the parametric amplifier operated in the nondegenerate mode, 
and $L$ is the small loss in the chain of the components between the output of the parametric 
amplifier and the input of the HEMT amplifier. 
Here, $T_{\rm Ni}$ should be almost equal to that of the vacuum noise 
$hf_{\rm s}/2k_{\rm B}=0.26$~K, and $G_{\rm nd} \ge 10^{1.4-0.3}$, 
where the factor $10^{-0.3}$ is due to the fact that 
the gain in the degenerate mode can be 3 dB larger than in the nondegenerate mode.~\cite{Yurke06} 
We estimate $L$ to be less than $10^{0.2}$. 
Consequently, the upper bound of $T_{\rm N}$ is estimated to be 0.87~K. 
We are planning more precise measurements using calibrated noise sources. 

In summary, we have designed and operated a flux-driven Josephson parametric amplifier. 
The band center of the amplifier is demonstrated to be widely tunable. 
We operated it at different band centers and 
observed amplification and deamplification depending on the relative phase between the pump and the signal. 
The noise temperature is estimated to be less than 0.87 K. 

The authors would like to thank Y. Yamamoto for fruitful discussions. 
They also would like to thank O. Astafiev, V. Bolkhovsky, and E. Macedo for technical assistance.

%\bibliography{squbit3}

\begin{thebibliography}{15}
\expandafter\ifx\csname natexlab\endcsname\relax\def\natexlab#1{#1}\fi
\expandafter\ifx\csname bibnamefont\endcsname\relax
  \def\bibnamefont#1{#1}\fi
\expandafter\ifx\csname bibfnamefont\endcsname\relax
  \def\bibfnamefont#1{#1}\fi
\expandafter\ifx\csname citenamefont\endcsname\relax
  \def\citenamefont#1{#1}\fi
\expandafter\ifx\csname url\endcsname\relax
  \def\url#1{\texttt{#1}}\fi
\expandafter\ifx\csname urlprefix\endcsname\relax\def\urlprefix{URL }\fi
\providecommand{\bibinfo}[2]{#2}
\providecommand{\eprint}[2][]{\url{#2}}

\bibitem[{\citenamefont{Takahashi}(1965)}]{Takahashi65}
\bibinfo{author}{\bibfnamefont{H.}~\bibnamefont{Takahashi}}, in
  \emph{\bibinfo{booktitle}{Advances in Communication Systems}}, edited by
  \bibinfo{editor}{\bibfnamefont{A.~V.} \bibnamefont{Balakrishnan}}
  (\bibinfo{publisher}{Academic, New York}, \bibinfo{year}{1965}), p.
  \bibinfo{pages}{227}.

\bibitem[{\citenamefont{Caves}(1982)}]{Caves82}
\bibinfo{author}{\bibfnamefont{C.~M.} \bibnamefont{Caves}},
  \bibinfo{journal}{Phys.\ Rev.\ D} \textbf{\bibinfo{volume}{26}},
  \bibinfo{pages}{1817} (\bibinfo{year}{1982}).

\bibitem[{\citenamefont{Barone and Paterno}(1982)}]{Baronebook}
\bibinfo{author}{\bibfnamefont{A.}~\bibnamefont{Barone}} \bibnamefont{and}
  \bibinfo{author}{\bibfnamefont{G.}~\bibnamefont{Paterno}},
  \emph{\bibinfo{title}{Physics and applications of the Josephson effect}}
  (\bibinfo{publisher}{Wiley}, \bibinfo{address}{New York},
  \bibinfo{year}{1982}), chap.~\bibinfo{chapter}{11}.

\bibitem[{\citenamefont{Movshovich et~al.}(1990)\citenamefont{Movshovich,
  Yurke, Kaminsky, Smith, Silver, Simon, and Schneider}}]{Movshovich90}
\bibinfo{author}{\bibfnamefont{R.}~\bibnamefont{Movshovich}},
  \bibinfo{author}{\bibfnamefont{B.}~\bibnamefont{Yurke}},
  \bibinfo{author}{\bibfnamefont{P.~G.} \bibnamefont{Kaminsky}},
  \bibinfo{author}{\bibfnamefont{A.~D.} \bibnamefont{Smith}},
  \bibinfo{author}{\bibfnamefont{A.~H.} \bibnamefont{Silver}},
  \bibinfo{author}{\bibfnamefont{R.~W.} \bibnamefont{Simon}}, \bibnamefont{and}
  \bibinfo{author}{\bibfnamefont{M.~V.} \bibnamefont{Schneider}},
  \bibinfo{journal}{Phys.\ Rev.\ Lett.} \textbf{\bibinfo{volume}{65}},
  \bibinfo{pages}{1419} (\bibinfo{year}{1990}).

\bibitem[{\citenamefont{Abdo et~al.}(2006)\citenamefont{Abdo, Segev,
  Shtempluck, and Buks}}]{Abdo06}
\bibinfo{author}{\bibfnamefont{B.}~\bibnamefont{Abdo}},
  \bibinfo{author}{\bibfnamefont{E.}~\bibnamefont{Segev}},
  \bibinfo{author}{\bibfnamefont{O.}~\bibnamefont{Shtempluck}},
  \bibnamefont{and} \bibinfo{author}{\bibfnamefont{E.}~\bibnamefont{Buks}},
  \bibinfo{journal}{Appl.\ Phys.\ Lett.} \textbf{\bibinfo{volume}{88}},
  \bibinfo{pages}{022508} (\bibinfo{year}{2006}).

\bibitem[{\citenamefont{Thol{\'e}n et~al.}(2007)\citenamefont{Thol{\'e}n,
  Erg{\"u}l, Doherty, Weber, Gr{\'e}gis, and Haviland}}]{Tholen07}
\bibinfo{author}{\bibfnamefont{E.~A.} \bibnamefont{Thol{\'e}n}},
  \bibinfo{author}{\bibfnamefont{A.}~\bibnamefont{Erg{\"u}l}},
  \bibinfo{author}{\bibfnamefont{E.~M.} \bibnamefont{Doherty}},
  \bibinfo{author}{\bibfnamefont{F.~M.} \bibnamefont{Weber}},
  \bibinfo{author}{\bibfnamefont{F.}~\bibnamefont{Gr{\'e}gis}},
  \bibnamefont{and} \bibinfo{author}{\bibfnamefont{D.~B.}
  \bibnamefont{Haviland}}, \bibinfo{journal}{Appl.\ Phys.\ Lett.}
  \textbf{\bibinfo{volume}{90}}, \bibinfo{pages}{253509}
  (\bibinfo{year}{2007}).

\bibitem[{\citenamefont{Castellanos-Beltran and Lehnert}(2007)}]{Beltran07}
\bibinfo{author}{\bibfnamefont{M.~A.} \bibnamefont{Castellanos-Beltran}}
  \bibnamefont{and} \bibinfo{author}{\bibfnamefont{K.~W.}
  \bibnamefont{Lehnert}}, \bibinfo{journal}{Appl.\ Phys.\ Lett.}
  \textbf{\bibinfo{volume}{91}}, \bibinfo{pages}{083509}
  (\bibinfo{year}{2007}).

\bibitem[{\citenamefont{Nakamura et~al.}(1999)\citenamefont{Nakamura, Pashkin,
  and Tsai}}]{Nakamura99}
\bibinfo{author}{\bibfnamefont{Y.}~\bibnamefont{Nakamura}},
  \bibinfo{author}{\bibfnamefont{Y.~A.} \bibnamefont{Pashkin}},
  \bibnamefont{and} \bibinfo{author}{\bibfnamefont{J.~S.} \bibnamefont{Tsai}},
  \bibinfo{journal}{Nature} \textbf{\bibinfo{volume}{398}},
  \bibinfo{pages}{786} (\bibinfo{year}{1999}).

\bibitem[{\citenamefont{Wallraff et~al.}(2004)\citenamefont{Wallraff, Schuster,
  Blais, Frunzio, Huang, Majer, Kumar, Girvin, and Schoelkopf}}]{Wallraff04}
\bibinfo{author}{\bibfnamefont{A.}~\bibnamefont{Wallraff}},
  \bibinfo{author}{\bibfnamefont{D.~I.} \bibnamefont{Schuster}},
  \bibinfo{author}{\bibfnamefont{A.}~\bibnamefont{Blais}},
  \bibinfo{author}{\bibfnamefont{L.}~\bibnamefont{Frunzio}},
  \bibinfo{author}{\bibfnamefont{R.-S.} \bibnamefont{Huang}},
  \bibinfo{author}{\bibfnamefont{J.}~\bibnamefont{Majer}},
  \bibinfo{author}{\bibfnamefont{S.}~\bibnamefont{Kumar}},
  \bibinfo{author}{\bibfnamefont{S.~M.} \bibnamefont{Girvin}},
  \bibnamefont{and} \bibinfo{author}{\bibfnamefont{R.~J.}
  \bibnamefont{Schoelkopf}}, \bibinfo{journal}{Nature}
  \textbf{\bibinfo{volume}{431}}, \bibinfo{pages}{162} (\bibinfo{year}{2004}).

\bibitem[{\citenamefont{Ojanen and Salo}(2007)}]{Ojanen07}
\bibinfo{author}{\bibfnamefont{T.}~\bibnamefont{Ojanen}} \bibnamefont{and}
  \bibinfo{author}{\bibfnamefont{J.}~\bibnamefont{Salo}},
  \bibinfo{journal}{Phys.\ Rev.\ B} \textbf{\bibinfo{volume}{75}},
  \bibinfo{pages}{184508} (\bibinfo{year}{2007}).

\bibitem[{com({\natexlab{a}})}]{comment1}
\bibinfo{note}{In the current-pumping scheme, the pump modulates Josephson
  inductance only quadratically unless the SQUID is biased by a dc current.
  Therefore, in the degenerate-mode operation, the frequencies of the pump and
  the signal are equal.\cite{Beltran07}}

\bibitem[{\citenamefont{Wallquist et~al.}(2006)\citenamefont{Wallquist,
  Shumeiko, and Wendin}}]{Wallquist06}
\bibinfo{author}{\bibfnamefont{M.}~\bibnamefont{Wallquist}},
  \bibinfo{author}{\bibfnamefont{V.~S.} \bibnamefont{Shumeiko}},
  \bibnamefont{and} \bibinfo{author}{\bibfnamefont{G.}~\bibnamefont{Wendin}},
  \bibinfo{journal}{Phys.\ Rev.\ B} \textbf{\bibinfo{volume}{74}},
  \bibinfo{pages}{224506} (\bibinfo{year}{2006}).

\bibitem[{\citenamefont{Sandberg et~al.}()\citenamefont{Sandberg, Wilson,
  Persson, Bauch, Johansson, Shumeiko, Duty, and Delsing}}]{Sandberg08}
\bibinfo{author}{\bibfnamefont{M.}~\bibnamefont{Sandberg}},
  \bibinfo{author}{\bibfnamefont{C.~M.} \bibnamefont{Wilson}},
  \bibinfo{author}{\bibfnamefont{F.}~\bibnamefont{Persson}},
  \bibinfo{author}{\bibfnamefont{T.}~\bibnamefont{Bauch}},
  \bibinfo{author}{\bibfnamefont{G.}~\bibnamefont{Johansson}},
  \bibinfo{author}{\bibfnamefont{V.}~\bibnamefont{Shumeiko}}, 
  \bibinfo{author}{\bibfnamefont{T.}~\bibnamefont{Duty}}, \bibnamefont{and}
  \bibinfo{author}{\bibfnamefont{P.}~\bibnamefont{Delsing}},
  \bibinfo{journal}{Appl.\ Phys.\ Lett.}
  \textbf{\bibinfo{volume}{92}}, \bibinfo{pages}{203501}
  (\bibinfo{year}{2008}).

\bibitem[{com({\natexlab{b}})}]{comment2}
\bibinfo{note}{The quality factor is flux dependent. It tends to be smaller as
  we approach $\Phi/\Phi_0=0.5$. For example, it is 170 for
  $\Phi/\Phi_0=0.42$.}

\bibitem[{com({\natexlab{b}})}]{comment3}
\bibinfo{note}{The maximum gain increases as we increase $P_{\rm p}$. 
However, the device becomes unstable if we increase $P_{\rm p}$ too much, 
which prevented us from making the maximum gain much higher than 14 dB.}

\bibitem[{\citenamefont{Yurke et~al.}(1988)\citenamefont{Yurke, Kaminsky,
  Miller, Whittaker, Smith, Silver, and Simon}}]{Yurke88}
\bibinfo{author}{\bibfnamefont{B.}~\bibnamefont{Yurke}},
  \bibinfo{author}{\bibfnamefont{P.~G.} \bibnamefont{Kaminsky}},
  \bibinfo{author}{\bibfnamefont{R.~E.} \bibnamefont{Miller}},
  \bibinfo{author}{\bibfnamefont{E.~A.} \bibnamefont{Whittaker}},
  \bibinfo{author}{\bibfnamefont{A.~D.} \bibnamefont{Smith}},
  \bibinfo{author}{\bibfnamefont{A.~H.} \bibnamefont{Silver}},
  \bibnamefont{and} \bibinfo{author}{\bibfnamefont{R.~W.} \bibnamefont{Simon}},
  \bibinfo{journal}{Phys.\ Rev.\ Lett.} \textbf{\bibinfo{volume}{60}},
  \bibinfo{pages}{764} (\bibinfo{year}{1988}).

\bibitem[{\citenamefont{Yurke and Buks}(2006)}]{Yurke06}
\bibinfo{author}{\bibfnamefont{B.}~\bibnamefont{Yurke}} \bibnamefont{and}
  \bibinfo{author}{\bibfnamefont{E.}~\bibnamefont{Buks}}, \bibinfo{journal}{J.\
  Lightwave\ Tech.} \textbf{\bibinfo{volume}{24}}, \bibinfo{pages}{5054}
  (\bibinfo{year}{2006}).

\end{thebibliography}

\end{document}